# Weather regimes linked to daily precipitation anomalies in Northern Chile

Oliver Meseguer-Ruiz[a]*, Nicola Cortesi[b], Jose A. Guijarro[c], Pablo Sarricolea[d]

[a] Departamento de Ciencias Históricas y Geográficas, Universidad de Tarapacá
18 de Septiembre 2222, Arica, Chile
omeseguer@academicos.uta.cl

[b] Earth Science Department, Barcelona Supercomputing Center (BSC-CNS), Spain
nicola.cortesi@bsc.es

[c] State Meteorological Agency (AEMET), Balearic Islands Office, Spain
jguijarrop@aemet.es

[d] Department of Geography, University of Chile
psarricolea@uchilefau.cl

* Corresponding author

**Abstract**

Northern Chile is one of the most arid regions in the world, with precipitation mainly occurring during austral summer, between December and April 1966-2015. The aim of this study is to classify the main weather regimes derived from sea level pressure, surface wind speed, 500 or 250 hPa geopotential heights, in order to measure their influence on precipitation anomalies and determine if they can be considered sources of predictability of rainfall in this region. Four weather regimes were found to optimally describe atmospheric circulation in the study area and for each of the four levels described above. Using daily precipitation data from a network of 161 meteorological stations across the region, the rainfall anomalies associated with each weather regime were quantified. They are coherent with the direction of flow derived from pressure and geopotential anomalies, bringing humid air masses from the Amazon Basin or the Pacific. The transitions between the different regimes are also coherent, representing transitions to and from similar regimes. A few negative and significant trends in the persistence of different regimes were detected, most likely linked to the absence of anthropogenic warming in the Antarctic as opposed to the Arctic. Finally, two of the regimes derived from surface wind speed exhibit a negative and significant trend in its frequency of occurrence, determining a precipitation decrease in the south of the study area (28-30ºS), which can be compared with the Megadrought experienced in central Chile.

**Highlights**

- WRs classified from 10-m wind speed determine part of the precipitation trend in northern Chile
- WR clustering is consistent between different atmospheric levels
- Significant changes in WRs persistence were identified

# 1. Introduction

The central Andes, and in particular northern Chile, are characterised by a very complex topography, with a high elevation gradient (Figure 1). The Andes mountain range clearly separates this region into a coastal area to the west under the influence of cold and dry winds from the Pacific Ocean (Houston and Hartley, 2003), and a continental area to the east influenced by warm and wet easterly winds blowing from the Amazon Basin (Garreaud, 2009). Rainfall behaviour in this area is influenced by these contrasting air masses, with very strong variations in annual totals (Figure 1), but also a high intra-annual and inter-seasonal variability (Romero et al., 2013). Rainfall also vary depending on the forcing induced by the topography, with wet events in the eastern part of the study area derived from the Amazon Basin, and wet events in the west derived from the Pacific Ocean. Water availability and management is a key topic for regional policymakers in the region (Sarricolea and Romero, 2015).

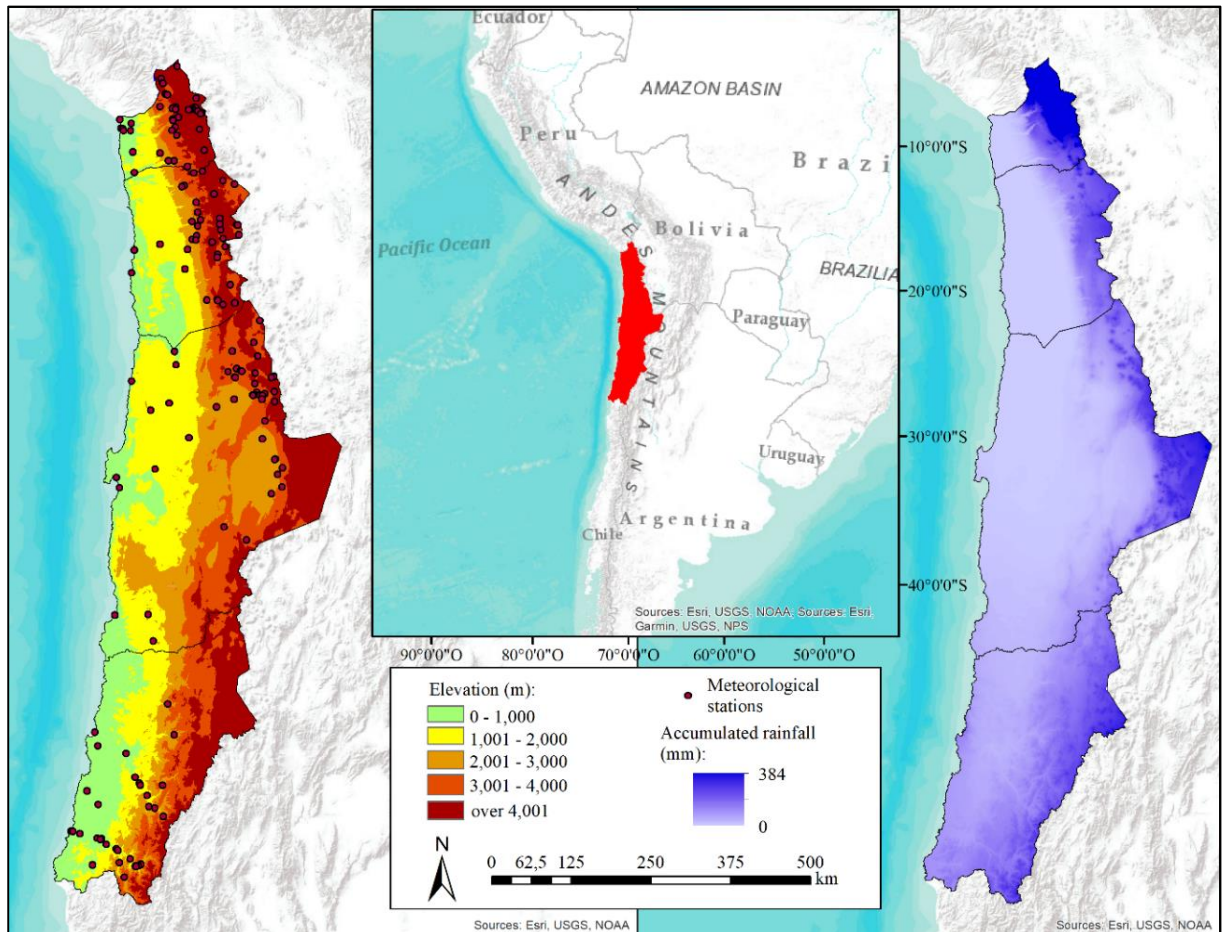

Figure 1. Map showing the topography of the study area and locations of the 161 meteorological stations employed (left), and the spatial distribution of mean annual precipitation (right) between 1950 and 2000 (Pliscoff et al., 2014).

Many studies have focussed on the mechanisms behind the spatial variation in precipitation in this region (Falvey and Garreaud, 2005; Valdés-Pineda et al., 2018; Candeo-Rosso et al., 2019). The El Niño-Southern Oscillation (ENSO) has been identified as the main regulator of precipitation (Garreaud and Aceituno, 2001; Junquas et al., 2013; Valdés-Pineda et al., 2015) on both annual and seasonal scales. During summer, an increase in precipitation is explained by an increase in vertical air instability and convection resulting from increased solar radiation, enhanced by the topography (Junquas et al., 2016, 2018). At upper levels (250 to 200 hPa), the Bolivian High, related to the South American monsoon, introduces a humid easterly air mass from the Amazon Basin, exhibiting a decadal and inter-decadal variability (Segura et al., 2016). A recent study (Zappalá et al., 2018) has linked permanent changes in precipitation patterns to a northward shift of the inter-tropical convergence zone (ITCZ). Moreover, an important decrease in precipitation (up to -50%) has been identified for central Chile (30°S-39°S) between 2010 and 2015, the so-called Megadrought (Boisier et al. 2016; Garreaud et al. 2017). According to these results, this event may also be evidenced in the south of northern Chile, around 29°S.
Climate change will have the most significant impact on arid regions (Donat et al., 2016), due to the higher sensitivity of natural processes and economic systems to variations in precipitation. Warming trends detected in the study area from 17ºS to 29ºS (Meseguer-Ruiz et al., 2018) can modify the hydrological cycle (Held and Soden, 2006), enhancing periods of drought (Sarricolea et al., 2017) and modifying the inter-annual and daily variability of precipitation (Sarricolea et al., 2019). Various factors, including the development of mining activity and increase in urban population, are leading to rising water demands in the area, based on the development strategy proposed by the country for the 2017–2022 period (Chilean Government, 2015). However, this development strategy ignores potential future water supply problems, as although precipitation projections show consistency on large spatial scales, they display significant uncertainty on smaller scales (Power et al., 2012). Kirtman et al. (2013) state that the projections are inconsistent in high and dry areas, such as within the central Andes. Cowtan and Way (2014) suggest that this could be related to the lack of observational data, which can lead to incorrect conclusions. Improving the quality and density of meteorological observations is an essential prerequisite for improving climate projections and, therefore, future water management.

Models projecting increasing precipitation in the north of Chile (CRUTS 4.02, dataset produced by the Climate Research Unit (CRU), University of East Anglia, Norwich, UK) do not differentiate natural variability from anthropogenic forcing. These projections exhibit changes which are magnitudes lower than the intrinsic error of the models (Hartmann et al., 2013), which is explained by the lack of observations in the study area (Trenberth et al., 2014). Studies focussed on the relationship between regional rainfall and large-scale forcing factors are important for understanding long-term variability, and for the development of effective future climate predictions. However, such works need to be complemented with studies of the relationship between rainfall and synoptic weather patterns at different tropospheric levels, which characterise the local features associated with rain days and dry days.

Regional-scale daily weather patterns which characterise the atmospheric circulation can be classified into discrete and recurrent categories, termed weather regimes (WRs), and are easy to simulate (Dawson et al., 2012; Dawson and Palmer, 2015). Synoptic WRs have been widely applied in multiple studies in the Northern Hemisphere (Yiou and Nogaj, 2004; Hertig and Jacobeit, 2014; Neal et al., 2016; Roller et al., 2016; Fereday, 2017; Cortesi et al., 2019). These show, in many cases, a correspondence with large-scale teleconnection patterns, such as the North Atlantic Oscillation (NAO) (Raymond et al., 2018), and important links with regional climate anomalies, including temperature and precipitation. WRs are usually defined by classifying a variable that helps to characterise the tropospheric circulation, such as the geopotential height, sea level pressure (SLP), and/or wind field (Schneider et al., 2018; Wang et al., 2019; Zhang and Villarini, 2019). For a recent characterisation of the uncertainty affecting near-surface wind speed from different analyses, see Torralba et al. (2017). Most of the studies on WRs are focussed on the winter season (Fil and Dubus, 2005; Ferranti and Corti, 2011; Vrac et al., 2014), or on the whole cold season from October to April (Ferranti et al., 2015), because WRs are more stable in time (more persistent) during this period and have a stronger influence on local climate (Cassou et al., 2005; Santos et al., 2005; Archambault et al., 2008). Changes in WR monthly frequencies of occurrence in the extratropics determine the majority of the variation of monthly precipitation anomalies (Beck et al., 2007), except in regions or periods in which large-scale circulation is less related to local precipitation, such as areas with complex topography (Salameh et al., 2008) or during the summer season in the boreal Hemisphere (Dünkeloth and Jacobeith, 2003). Because tropical circulation is dominated by other patterns (as explained) that are inherently more important than in the extratropics, the dynamic interpretations of WRs are less clear than in extratropical circulation situations, but can still be applied (Vigaud and Robertson, 2017). Nevertheless, other methods can also be applied while characterising concrete atmospheric dynamics at inter-tropical latitudes (Briceño-Zuluaga et al., 2017).

In the Southern Hemisphere, WR studies have been mainly applied to Australia and the western Pacific (Silvestri and Vera, 2003; Risbey et al., 2009; Wilson et al., 2013; Lorrey and Fauchereau, 2018). These studies have shown that some WRs were associated with moist/wet conditions and others with dry conditions. The WRs were phase-locked to the seasonal cycle with a moderate degree of coherency. Correlations between those WRs and global precipitation reanalysis fields indicate a strong connection between regional weather patterns and the South Pacific Convergence Zone (SPCZ), whilst the South Atlantic Convergence Zone (SACZ) controlled the position and relative intensity of convective activity within the SPCZ (Zilli et al., 2019). There are significant impacts from south-west Pacific WR frequency changes and regime persistence on extreme rainfall deficits and/or surpluses for small areas during the summer. Another WR is connected to an enhancement of the Hadley-Ferrell circulation, while two other WRs are influenced by low-frequency teleconnection patterns, such as the El Niño-Southern Oscillation (ENSO), the Madden-Julian Oscillation (MJO), or the Antarctic Oscillation (AAO). Lastly, we conclude that WR investigations can help to understand spatial-scale mismatches that exist between global and regional models for areas characterised by a low density of meteorological observations in the context of future climate change scenarios.

Similar studies were also carried out in North America (Robertson and Ghil, 1999; Sheridan, 2002; Casola and Wallace, 2007; Yu et al., 2015; Roller et al., 2016; Díaz-Esteban and Raga, 2018) and Central America (Sáenz and Durán-Quesada, 2015). The intra-seasonal variability of the WRs and their associated rainfall are well captured by these studies. WRs exhibit negative SLP anomalies, southerly winds, above-average low-level moisture, and little influence of the North Atlantic Subtropical High, which results in above-average precipitation. ENSO phases are also clearly identified in the low-latitude studies.

However, WR studies applied to the South American continent are even fewer, being limited to south of 35ºS (Solman and Menéndez, 2003). A robust classification of five WRs was defined from the most frequent large-scale circulation anomalies. The most significant change in precipitation frequency is detected for a ridge over the south-east Pacific and NW-SE trough downstream, with wetter conditions over all of the analysed regions. A recent study (Zappalà et al., 2018) has related significant changes in precipitation and ice melting to a northward shift of the ITCZ, and to a widening of the rainfall band in the western Pacific Ocean (ascending branch of the Hadley Cell), using the Hilbert amplitude.

The present study aims to determine the influence of WRs on precipitation anomalies for the rainy season, from December to April, and its implications regarding persistence and transitions in northern Chile. Summer (winter) is the rainy (dry) season in this area, while it is a dry (wet) season in the rest of the country. This should allow identifying the synoptic patterns that are related to precipitation in the study region. Three distinct tropospheric levels were analysed to identify different atmospheric processes, including at SLP, 500 hPa (g500), and 250 hPa (g250) geopotential heights, besides the WRs classified from the 10 m wind speed field. Moreover, a trend analysis was conducted in order to identify significant changes in the frequencies of WRs and possible effects on future precipitation.

This work is organised as follows. Firstly, section 2 describes the dataset quality control, the precipitation anomalies, and the method applied to determine the WRs, their transitions, and persistence. Section 3 presents the main results of the study and their comparisons with other studies. Their discussion was drawn in section 4, along with the conclusions.

**2. Data and methodology**

2.1. Observed rainfall data

Daily rainfall records from 161 stations across the study area for the period 1 January 1966 to 31 December 2015 were used. Quality control was performed using the Climatol version 3.0 package in the R software (Guijarro, 2016), which uses normal ratio values (data are divided by the mean of its series) of the closest precipitation data to build reference series for all the stations. Differences between the observed and reference series were used to test their quality by outlier detection, and their homogeneity was assessed using the Standard Normal Homogeneity Test (SNHT) (Alexandersson, 1986). At the same time, undoing the normalisation of the reference series provides estimations to fill any missing data in the series. The detection of significant shifts in the mean was done on the monthly aggregates of the series, since the much higher variability of the daily series makes the detection of variations in the mean more difficult (Vincent et al., 2002), especially in arid climates. The database has already been used in a previous study (Meseguer-Ruiz et al., 2019), and its quality has been tested and accepted.

2.2. Daily anomalies

Daily anomalies for precipitation, SLP, geopotential heights, and wind speed were computed from the daily climatology of the respectives datasets (explained in sections 2.1 and 2.4) for the period 1966–2015, previously filtered with a local polynomial regression (LOESS), to take into account the annual cycle and smooth the short-term variability (Mahlstein et al., 2015). Before classifying the WRs, daily gridded SLP and geopotential heights anomalies were weighted by the cosine of the latitude, in order to ensure equal area weighting at each grid point.

2.3. Precipitation trend

To determine the evolution of the accumulated rainfall during the wet season (from December to April) between 1966 and 2015 in the studied area, we applied the Mann-Kendall (MK) non-parametric test (Mann, 1945; Kendall, 1962) with a 95% significance level. Trend was expressed in percentage, to take into account the great differences of the total rainfall amount between stations in the study domain, as shown in Figure 1.

2.4. Weather regimes classification

Global National Centers for Environmental Prediction (NCEP)/NCAR Reanalysis v1 product (Kalnay et al., 1996) was used to determine the different WRs between 1966 and 2015. This product represents one of the most complete and physically consistent atmospheric datasets for the Southern Hemisphere, and it can be considered to be reasonably homogeneous for the study of large-scale circulation during this period. WR classifications for South America are not frequent (Solman and Menéndez, 2003), but other approaches have been carried out with synoptic weather types (Sarricolea et al., 2014, 2018), following the Jenkinson and Collison classification. In order to link the WRs with regional climate, precipitation station data were utilized, as explained in section 2.1.

To identify optimal relationships, four different WR classifications were considered, each one derived by a distinct predictor: SLP, surface wind speed, g500 or g250 heights. The predictors were not combined in a single classification because this is the first time regimes are classified in Northern Chile, so we wanted firstly to analyse the differences between the four regime classifications, in terms of different regime patterns, frequencies, persistences and transition probabilities. Future work might focus on a single classification that combines two or more predictors. SLP and wind speed present important trends during 1966-2015 over part of South America (Figure 2), while g500 and g250 present trascurable ones, limited to a few meters. Usually, to remove long-term bias in the WR daily time series, prior to the *k-means* clustering the long-term linear trend of the circulation variables is removed from the original data, along with the precipitation trend (Lorrey and Fauchereau, 2018). However, in this case, given the trends shown in Figure 2 and 4, it is also interesting to analyse to which extent these trends influence each of the four WR classifications. Thus, WRs were also classified without detrending any variable; results for these alternative WR classifications were shown in the supplementary material and discussed in Section 3.

A *k-means* cluster analysis was applied to the predictor's daily anomalies to classify four WRs (Hartigan and Wong, 1979; Hart et al., 2006) from December to April of 1966–2015, employing the methodology of Cortesi et al. (2019), albeit for a different spatial domain (16.25–31.25°S, 63.75–76.25°W) and period (DJFMA 1966-2015). In short, each of the four predictors is firstly converted to daily anomalies with a LOESS polynomial filter to remove seasonal cycle. Then, anomalies are weighted for latitude by multiplying them for the cosine of the latitude, in order to ensure equal area weighting of each grid point prior to the clustering. Finally, the *k-means* algorithm is applied to the anomaly series of DJFMA 1966-2015.

The spatial domain of the cluster analysis is also shown as a small gray box in Figures 4-7; it is much smaller than the typical one employed in literature for classifying WRs in the extratropics, because in the tropics SLP and geopotential varies much less than in the extratropics. As a consequence, the clustering domain cannot include both tropical and extra-tropical regions (as for example the region shown in Figure 2), or the WR patterns would be mostly determined by variations in the extra-tropical part of the domain, generating null WR pattern anomalies in the tropical part of the domain for all WRs. For this reason, a small domain was chosen, including only the northern part of Chile, where both SLP and geopotential varies homogeneously through the region.

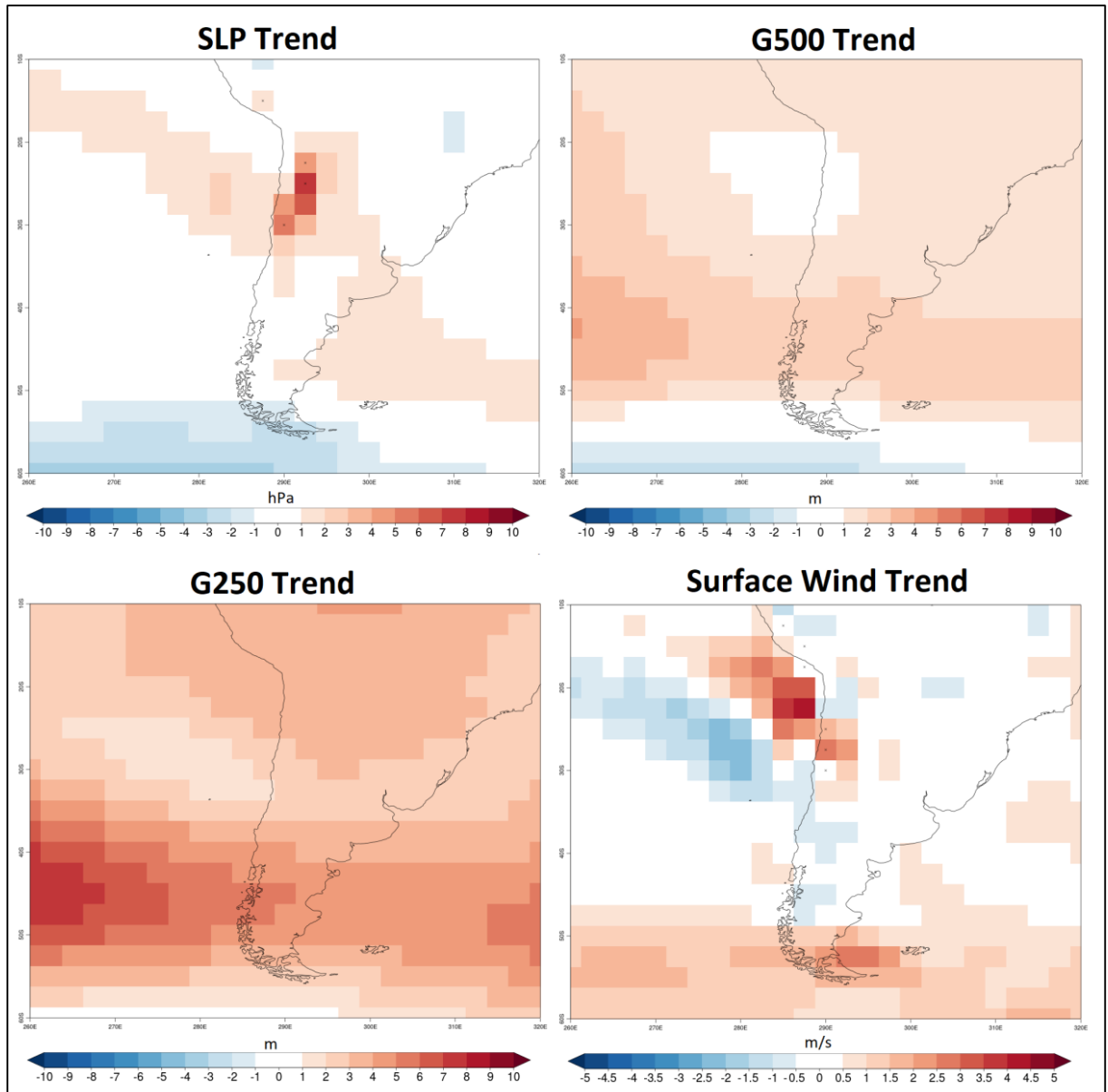

Figure 2. DJFMA trends of SLP (top left), g500 (top right), g250 (bottom left), and wind speed (bottom right) for the whole period 1966–2015. Source: National Centers for Environmental Prediction (NCEP) v1 reanalysis.

The main caveat of the *k-means* is that the optimal number $N$ of clusters is not defined *a priori*, so a criterion needs to be introduced. In this work, 23 different objective criteria for determining the optimal number of clusters were tested (Charrad et al., 2015), and the distribution of the number of criteria as a function of the number of clusters used to classify the WRs was plotted, to determine the optimal number of clusters (Figure 3). Classifications with three or four clusters are recommended by the highest number of criteria, therefore a classification based on four clusters was chosen in this work, to increase the probability of finding at least one regime strongly linked with precipitation. Cluster analysis associates each day to one of the four chosen WRs based on its normalised distance from the four cluster centres, leaving no days unclassified. Lastly, the frequency trend for each WR was measured using the Mann-Kendall (MK) non-parametric test (Mann, 1945; Kendall, 1962) with a 95% significance level.

Several tests were developed to identify the optimal spatial domain to classify the WRs for northern Chile. The WR classification showed a very high sensitivity to the spatial domain, as WRs defined over a small domain do not show any clear patterns or anomalies. On the other hand, if a bigger domain is employed, the influence of WRs on the precipitation decreases.

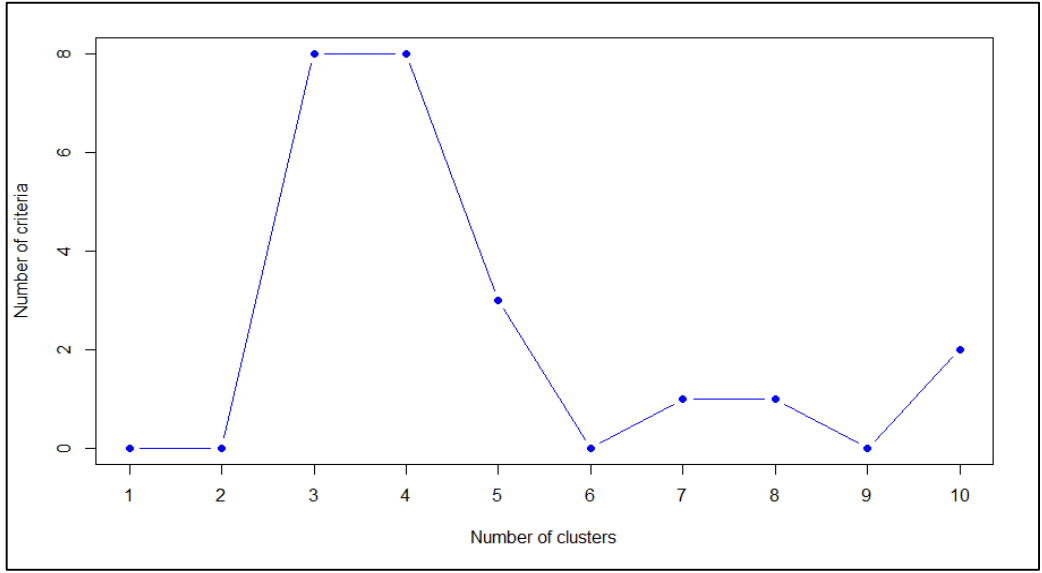

Figure 3. Number of different criteria for SLP, as a function of the number of clusters of the *k-means*. Other variables (g500 height, g250 height, and 10 m wind speed) exhibit similar characteristics. The *k-means* spatial domain for northern Chile is shown in Fig. 4.

The WRs were also linked to NCEP/NCAR daily precipitation reanalysis data (not shown), but its spatial resolution (2.5°) was not suitable for the aims of the project. Furthermore, a comparison between the reanalysis and the observed data revealed many critical differences for many of the meteorological stations. This is mainly due to the geographic characteristics of the study region (as shown in Fig. 1). This complex orography, with very significant differences of elevation and orientation in very few kilometres, is related to a high spatial variability of precipitation (Meseguer-Ruiz et al., 2019), with stations with very few accumulated rainfall close to the coast, and stations with accumulated rainfall higher than 300 mm in high elevations.

Finally, the exact amount of long-term rainfall changes induced by WRs at a particular station can be measured from the frequency change of both WRs and their impact on precipitation at each station. The change is given by the linear combination of the impact on precipitation of each WR for its frequency change:

$$\Delta Prec = \sum_{wr=1}^{4}(I_{wr} \cdot \Delta\omega_{wr}) \tag{1}$$

being $I_{wr}$ the impact of a WR on daily rainfall (in %) and $\Delta\omega_{wr}$ its frequency change (in %) from 1966 to 2015.

To derive equation (1), we start noticing that the impact of a regime on rainfall represents the average daily precipitation anomaly (in %) associated to the regime (as shown in the left panels of figures 5-8). Thus, a value of 30% at a particular station means that on a day belonging to that regime, precipitation at that point is on average 30% higher than its climatological value. If the same regime persists for a whole austral summer, the summer precipitation anomaly estimated would be exactly 30% higher than the climatological value, as all days of that period belong to the same regime. However, usually two or more regimes are observed during an austral summer. As a consequence, to measure the summer precipitation anomaly estimated by the regimes, the % impact on precipitation of a given regime is weighted by its frequency of occurrence (also in %) in that summer, and it is summed over the similar contributions of the other three regimes, obtaining an equation like (1) but without the two Δ. This relationship is valid for the present climate, but also for the past and future climate. So, if we measure summer precipitation in 1966 and in 2015, we can derive two similar equations, changing only the frequencies of the regimes, as the impact on regimes on precipitation is the same thorough all the study period 1966-2015. To get the precipitation change from 1966 to 2015 estimated by the regimes, we simply subtract the first equation from the second, obtaining equation (1).

Notice that in case of the WRs derived from surface wind speed (Figure 8), only two WRs, the first and the fourth, exhibit a not negligible frequency trend, so equation (1) can be simplified as:

$$\Delta Prec = I_1 \cdot \Delta\omega_1 + I_4 \cdot \Delta\omega_4 \tag{2}$$

2.5. WR transitions, persistence, and trends

Transitions between regimes are simply defined as the percentage of shifts from one WR to another. In this work, we have considered the transitions between events corresponding to two different WRs, *M* and *N* (with *M, N* = 1–4). These transitions are calculated as the number of times that a transition occurs from event *M* to event *N*, divided by the total number of the events. We only considered cases in which both events had durations of longer than four days. Shorter daily sequences were discarded to filter out noise, ensuring in this way that only real transitions between the two selected regimes are considered. The significance level for transitions considered was 95%.

Analysis of persistence (average of consecutive days belonging to the same WR) and transition among WRs is worth studying in terms of the linkage with local weather conditions, mainly over the continental area. In particular, persistence of anomalous conditions in local weather could be inferred from WR persistence and their transitions. WR persistence has been measured and its trends determined with the MK test (significance level = 95%).

### 3. Results and discussion
#### 3.1. Data quality
The Climatol homogenisation was applied to the precipitation monthly aggregates using a conservative approach and a threshold of SNHT = 20 for the detection of shifts, in order to avoid false positives. This procedure showed up 52 break points in the monthly aggregate series and were adjusted consequently. Daily series were then adjusted by splitting at those dates in order to reconstruct complete series from every homogeneous sub-period. However, due to the extreme aridity of most of the studied territory, most daily series have means of lower than 1 mm, jeopardising the normal ratio normalisation used in the process. Therefore, no adjustment was applied to the final daily series, which was only completed by infilling all of the missing values. Nonetheless, the impact of the correction would have been negligible, due to the extremely small amounts of precipitation recorded in the area. Only the closest reference data item at every time step was used in this process, to avoid the smoothing effect that would have resulted from the use of several reference data, which in an arid climate with isolated precipitation events, would have increased the number of rainy days at the expense of averaging the values of rainy and rainless spots in the vicinity. No outstanding outliers were identified through this process that could reliably be identified as errors, and therefore all original data were kept (as shown in a previous study, Meseguer-Ruiz et al., 2019).

#### 3.2. Precipitation trends
Generally, precipitation decreased in northern Chile during the wet season (DJFMA), as shown in Figure 4. Negative trends is especially high along the Pacific Coast and to the south of the area, with some stations exhibiting a decrease of -60% or lower. However, due to the very high temporal variability of daily rainfall in northern Chile, the trends are rarely significant. The only stations with significant trends are found in the northeastern part of the domain, where rainfall in less variable.

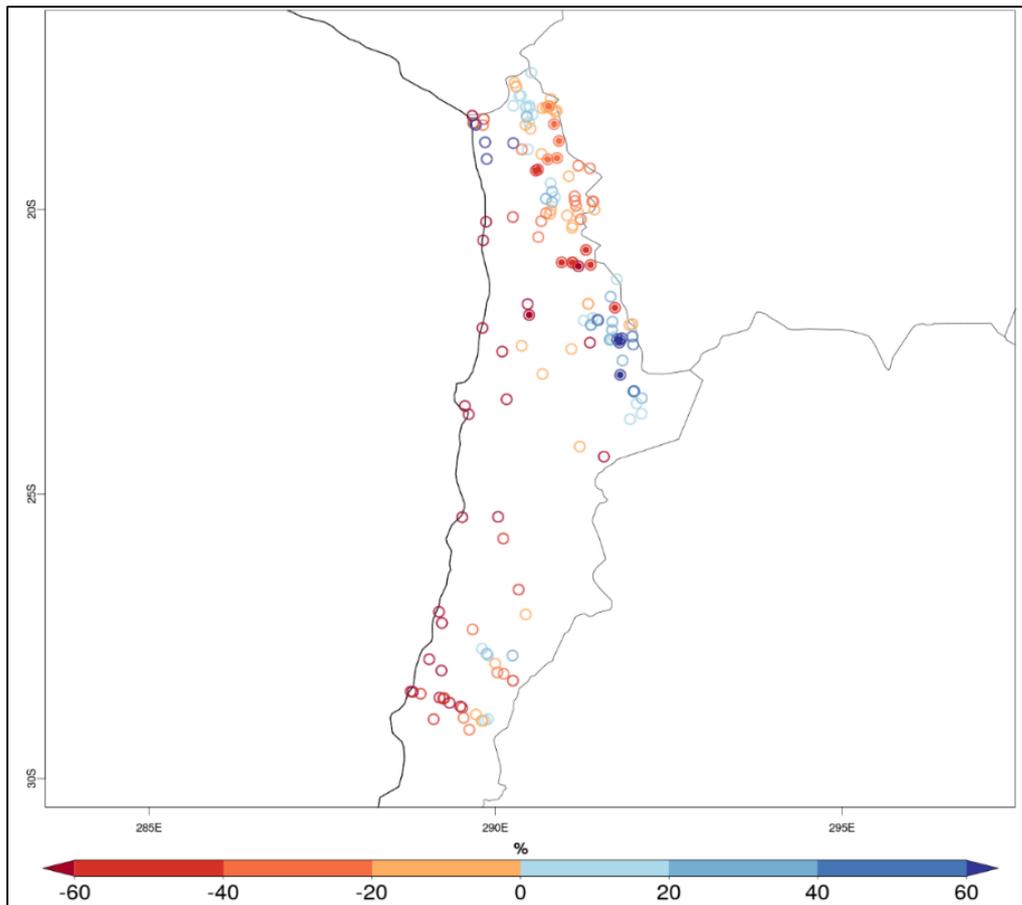

Figure 4. Precipitation trend (%) during the wet season (DJFMA) between 1966 and 2015. Eventual solid dots show significant trends for the MK test at the 95% confidence level.

3.3. WR patterns and their impacts on precipitation

In section 2.3, four WRs were selected to optimally describe the main weather patterns observed in northern Chile. Figure 5 shows, from top to bottom, the four WR patterns in decreasing order of average frequency 1966-2015 and classified during the wet season (December to April) from SLP (center column), along with their related precipitation anomalies at each meteorological station in the study area (left column) and their inter-annual frequency of occurrence (right column).

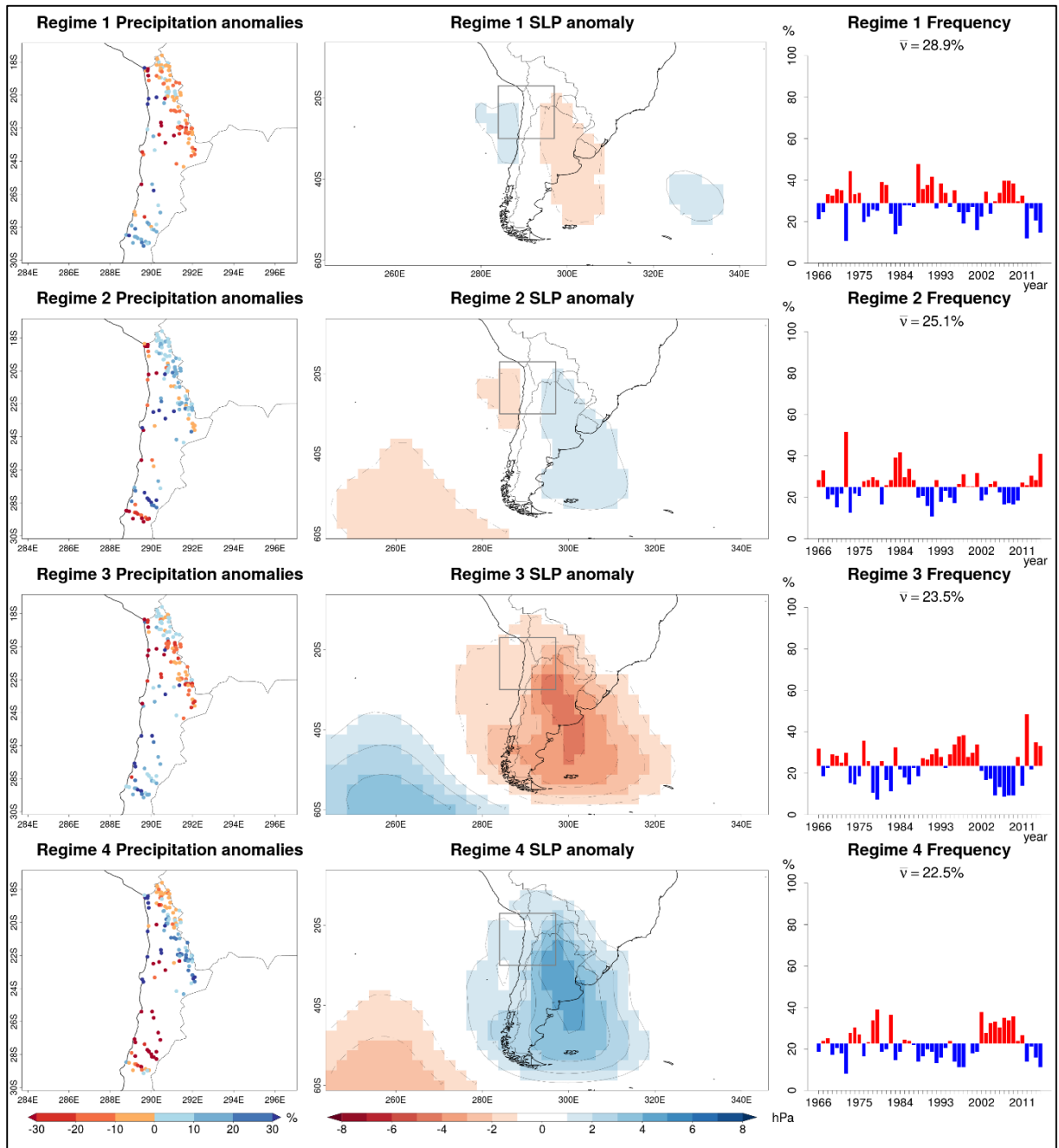

Figure 5. Precipitation anomalies (left, in %), sea level pressure anomalies (centre, in hPa), and inter-annual frequency (right, in %) associated to each of the four WRs between December and April for 1966–2015. WRs are ordered by decreasing frequency of occurrence from top to bottom. Station network over northern Chile shown to the left is the same highlighted in the grey frame at the center and it is also the spatial domain employed in the clustering analysis. Eventual black dotted lines to the right show significant trends for the Mann-Kendall test at the 95% confidence level. Source: NCEP v1 reanalysis.

At SLP, the spatial patterns of the first WR are characterised by weak negative SLP anomalies (up to -2 hPa) to the east of the Andes and weak positive ones (up to +2 hPa) to the west, with associated negative precipitation anomalies in the north of the study area (up to -30%) and positive anomalies in the south (up to +30%). Its average frequency of occurrence is 28.9%, the highest of the four WRs. This synoptic configuration is probably derived from the influence of the prevailing westerlies southward of 30ºS. The orographic effect of the Andes explains the positive precipitation anomaly in the southern part of the study region and the negative one in the north, where the influence of the South Pacific High is evident, blocking western advection.

The second WR represents a similar pattern to the first WR but with an opposite sign, with the positive anomaly located over southern Argentina and the southwestern Atlantic. This also determine a change in the sign of the precipitation anomalies, as positive ones are observed in elevated meteorological stations to the north of the study area, such as the Altiplano, while negative anomalies are recorded in stations the extreme south. Rainfall anomalies in the north are explained by the advection of humid air from the Amazon Basin. This north-easterly air mass is dry by the time it reaches the southern part of the study area due to the katabatic effect, explaining the negative precipitation anomaly in the southern stations.

The third WR shows an intense and negative SLP centroid over Argentina (up to -5 hPa) and a positive one (up to +4 hPa) located in the Pacific Ocean, to the south-west of the continent. Positive anomalies are present in stations located southwards, towards the Tropic of Capricorn while negative anomalies in the stations located in the north, except for those located northwards of 20ºS and at high altitude areas, over 3,000 m above mean sea level (ASL). This WR is characterized by a humid flow from the Pacific Ocean, associated with cold fronts that generate more precipitation in the south of the study area.

The fourth WR exhibits an opposite pattern to that shown by the third WR, including a positive centroid (up to +5 hPa) over South America, and a negative one over the south-west. This leads to negative precipitation anomalies in the south and in the northern Altiplano, and positive precipitation anomalies in elevated areas of the central region. The fourth WR also has the lowest frequency of occurrence (22.5%).

At the 500 hPa level, the first WR pattern shows almost no anomalies (Figure 6) and only a very small area with a positive anomaly (up to +1 m). On the contrary, precipitation exhibits a very contrasting pattern, with both positive and negative anomalies but no clear spatial gradient, except in the southern region, where precipitation anomalies are mostly positive. Its average frequency is very high, 36.3%, meaning that more than one day each three belongs to this WR. Thus, the more frequent synoptic situation at 500 hPa is associated to the average atmospheric circulation (null anomalies). For this reason, the precipitation anomalies associated to the first WR can be linked to the regional and local conditions derived from the topography (Canedo-Rosso et al., 2019).

The second WR is characterized by negative geopotential anomalies (up to -3 m) over the continent between 20ºS and 50ºS, and two weak positive anomalies (up to +2 m) located in the south-western Atlantic and in the south-eastern Pacific. Precipitation anomalies are positive in the elevated areas in the north, negative just to the north of the Tropic of Capricorn, and generally positive in the south, but negative in the extreme southern part of the study area. This synoptic configuration could be linked with the findings of Junquas et al. (2016, 2018), where precipitation changes are induced by changes in the convective processes. The same phenomenon was identified in topographically high areas in the world, such as the Tibetan Plateau (Ge et al., 2019).

The third WR has similar geopotential anomalies of the second one but more intense and of opposite sign: positive anomalies reach up to +5 m and are centred over central Chile and Argentina, while two negative centroids (up to -4 m) are present over the south-eastern Pacific and the south-western Atlantic. Precipitation anomalies are mainly negative in the inter-tropical Altiplano (except for stations located in the far north), but are both slightly positive and negative south of 23ºS.

The fourth WR is characterised by strong and negative (up to -7 m) anomalies over central Chile and Argentina and positive (up to +6 m) anomalies located over south-eastern Pacific, similarly to the patterns of the second WR but much more intense. Negative precipitation anomalies were observed in the north and south of the study area, with positive anomalies in the elevated parts of the central zone.

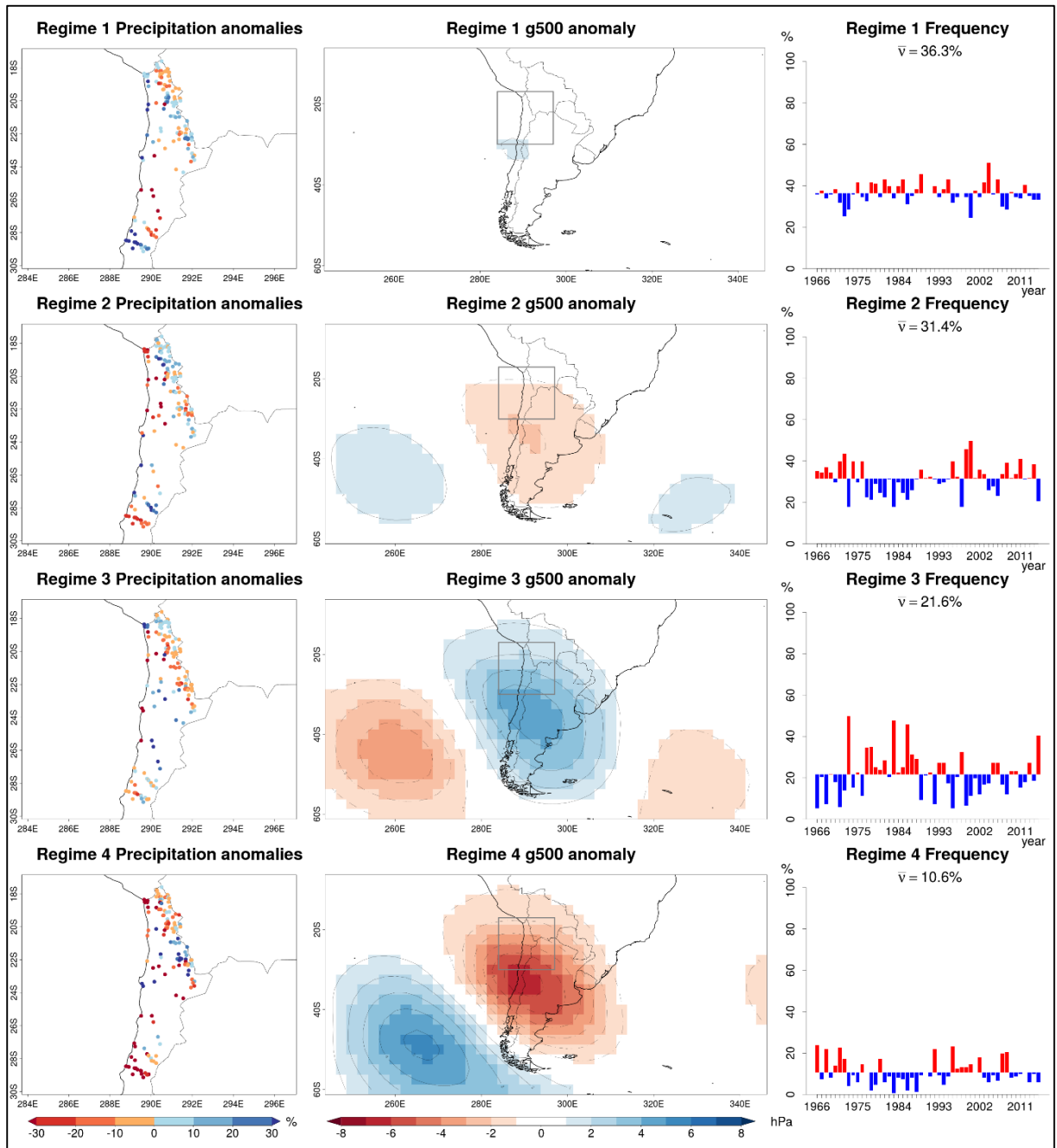

Figure 6. As Figure 5, but for geopotential anomalies at 500 hPa (in m).

It is interesting to notice that the third and fourth WRs show almost opposite anomalies, which is also evident at g250 heights (Figure 7). Thus, the high troposphere exhibits the same opposing anomalies as the mid troposphere (g500 height). In fact, at both g500 and g250 heights, the third WR forms a deep positive anomaly over the continent, expressing the presence of the Bolivian High in the mid, but most noticeably, the high troposphere. This agrees with a number of studies which suggest that formation of the Bolivian High brings the majority of precipitation to northern Chile (Sarricolea and Romero, 2015; Meseguer-Ruiz et al., 2019). Very intense radiation and heating (Meseguer-Ruiz et al., 2018) over elevated areas initiate convective processes, allowing air to accumulate in the mid to high troposphere, forming the Bolivian High. This high pressure brings moist air masses westward from the Amazon Basin, creating an influx of vast amounts of water vapour, resulting in positive precipitation anomalies in topographically high areas of northern Chile.

The opposite states for the fourth WR (for both g500 and g250 heights) are in agreement with Segura et al. (2016) and Zappalá et al. (2018), linking changes in wind circulation in the high troposphere (at 200 hPa) to hydroclimatic variability over the central Andes. This may affect the coastal branch of the subtropical anticyclone and generate anomalies in the Walker circulation. These modifications to atmospheric circulation alter the normal behaviour of the South American monsoon described by Sarricolea and Romero (2015), shifting the Bolivian High northward in the high troposphere.

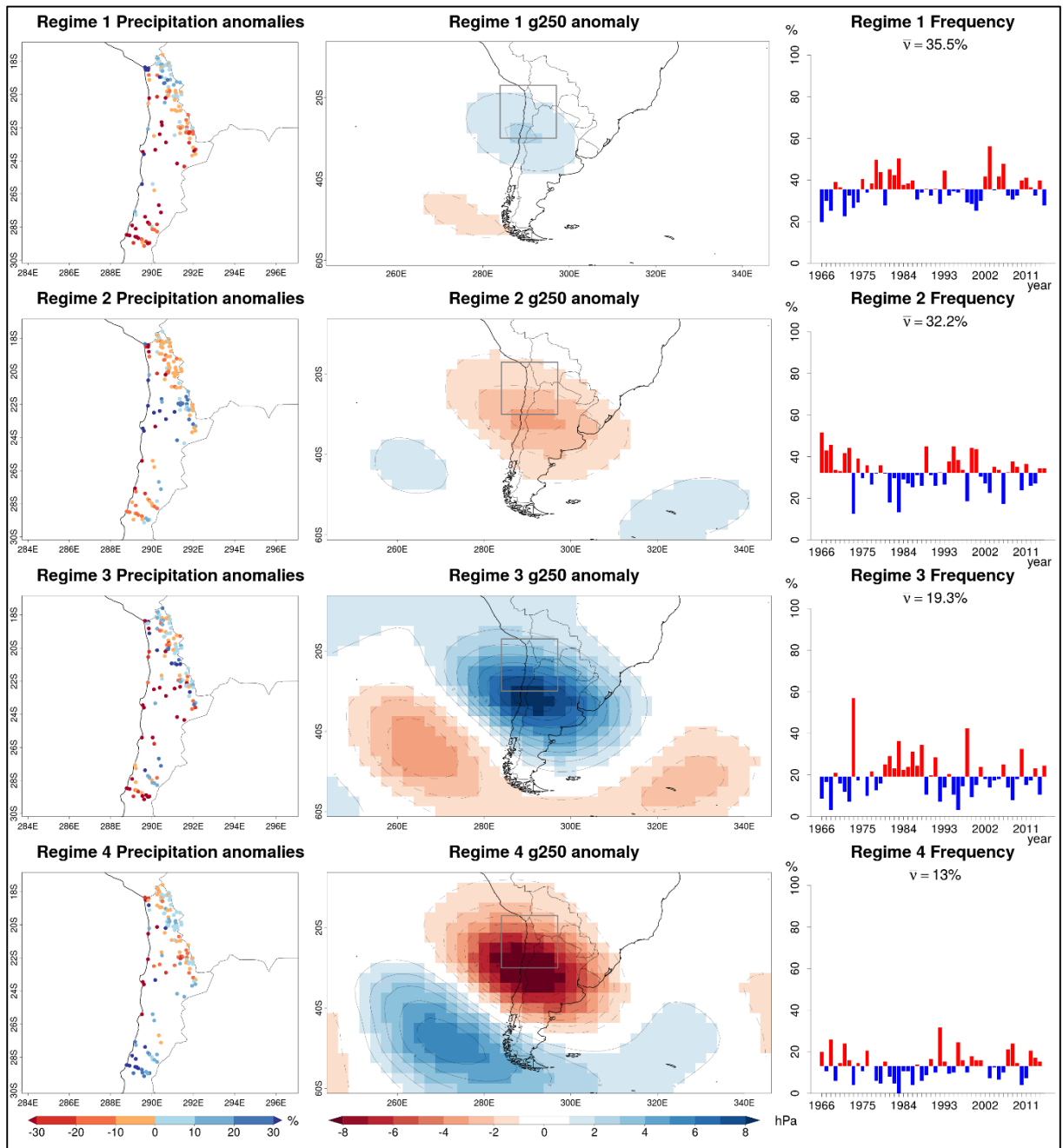

Figure 7. As Figure 5, but for geopotential anomalies at 250 hPa (in $m^2 \cdot s^{-2}$).

Figure 7 illustrates the four regimes at the 250 hPa level, and their frequency and impacts on observed precipitation patterns. The first WR represents a frequent configuration of the Bolivian High, centred over northern Chile. It is characterised by a positive geopotential anomaly (up to +3 m) above northern Chile and a weak negative anomaly (up to -2 m), with a very high average frequency of 35.5%. This pattern is linked to positive precipitation anomalies at stations located in the elevated areas of the north, but negative anomalies in topographically low areas of the north and in the central and southern parts of the study area.

The second WR shows a negative geopotential anomaly (-3 m) centred at 30ºS, and two weak positive ones (both up to +2 m) over the Pacific and Atlantic oceans. Precipitation anomalies are negative in the north of northern Chile and in the southern part of the area, but are positive in the central zone. This WR leads to a zonal circulation sourced from the ocean, with humid air masses arriving through the continent and configuring positive precipitation anomalies in the central part of the study area. In this case, the topography of Chile plays a key role (Junquas et al., 2018).

The third WR exhibits a large positive geopotential anomaly centred around 30°S (up to more than +8 m) and a negative one (up to -3 m) in the mid-latitudes between 20°W and 100°W. Precipitation anomalies are positive in elevated areas, but negative in low areas and in the extreme south of the area. This WR reproduces the same mechanism shown for the first WR, but with more moderate positive rainfall anomalies.

The fourth WR presents a pattern opposite to that of the third one, with strong and negative geopotential anomalies (up to more than -8 m) located above northern Chile and a positive anomaly (up to -6 m) centred west of the extreme south of South America. Its mean frequency is the lowest of the four WRs (only 13%), and it is linked to positive precipitation anomalies in the southern part of the study area, but it does not show any clear rainfall pattern in the central and northern areas.

Figure 8 illustrates the classification based on 10-m wind speed anomalies, along with the WR frequencies and impact on precipitation anomalies. The first WR has the highest frequency of occurrence (27.7%) and is characterized by negative wind anomaly (up to -1.5 m/s) over the Pacific coast of northern Chile, and by positive anomalies (up to +2 m/s) over the central coast of Chile. It is the only WR with a negative and significant trend during the study period. It negatively affects precipitation anomalies at stations located in the extreme north and positively affects anomalies at stations located in the extreme south.

The second WR present a positive centroid with wind anomalies up to +3 m/s over the tropical and subtropical Pacific coast of South America. It is related to significant negative precipitation anomalies in the south of the study region, but no clear rainfall patterns are identified elsewhere in the study area.

The third WR exhibit a single centroid as the previous WR, but of opposite sign, with negative wind anomalies of up to -3 m/s. It is mostly linked to positive precipitation anomalies, most evident in the extreme south, with only a few stations showing negative anomalies in the central zone at moderate elevations (from 1,500 to 2,500 m ASL). This might mean that precipitation occurrence is more dependent on wind direction (dry versus humid flow) rather than wind speed.

The fourth WR is characterised by the lowest frequency (23.6%) and shows a wind pattern similar to the first WR, but of opposite sign: negative wind anomalies (up to -2 m/s) are present over the tropical coast of the Pacific and positive anomalies (up to +2 m/s) are observed over the subtropical coast. This pattern determines dry conditions in the south of the study area, but no homogeneous impacts on precipitation were found in the central and northern parts of northern Chile.

It is worth noting that the negative and significant trend in the frequency of the first WR is mirrored by a positive trend of the frequency of the fourth WR, even if in the last case the trend is not significant. The other two WRs do not show any frequency change, so we can interpret these trends as if a fraction of the days belonging to the first WR were lost in favour of the fourth WR. Moreover, the first WR is associated to positive precipitation anomalies in the south, while the fourth WR is linked to negative ones in the south. For this reason, such a long-term change of the frequency of the first and fourth WR determines an overall decrease of precipitation in the south, up to 29° S.

These results can be compared with the findings of Boisier et al. (2016) and Garreaud et al. (2017), evidencing a significant decrease of the precipitation amount in the northern part of central Chile, part of the area interested by the Megadrought (30-38° S). These two WRs involved present the same surface wind patterns but with opposite sign, so in both cases the precipitation deficit can be attributed to a decrease in surface wind speed over the subtropical Pacific coast of Chile and an increase over the tropical coast. Thus, such a deficit can be related to weaker subtropical westerlies, which are also linked to the negative phase of the ENSO (Solman and Menéndez, 2003). Furthermore, even if the second and third WRs do not present any frequency trend, they still present a strong impact on precipitation in the south. WR influence on the central and northern part of the study domain is usually more heterogeneous, with no clear spatial patterns. The same consideration is also true for the WRs derived from SLP and geopotential.

The frequency of the first WR almost halved during 1966-2015 (from 35% to 17% of the total number of DJFMA days (see top right of Figure 8), while its impact on precipitation in the southern part of northern Chile is on average 25%, depending on the station (Figure 8, top left). Similarly, the frequency of the fourth WR increased of roughly 15% during the study period (Figure 8, bottom right), even though not significantly, while its impact on rainfall in the southern part of northern Chile is roughly 30% (Figure 8, bottom right). Thus, equation (2) becomes:

$$\Delta Prec = 25\% \cdot (-18\%) + (-30\%) \cdot 16\% = -4.5\% - 4.8\% \cong -10\% \qquad (3)$$

Thus, the two terms together determine an average rainfall decrease in the south of the domain of roughly -10%, smaller that the observed average decrease of -20% to -60% (Figure 4). This is partly because both precipitation and wind speed were detrended prior to the *k-means* clustering (see Section 2), so the four WR wind anomalies in Figure 8 refer to the detrended fields. Detrending is necessary for WRs to present interdaily variability: without detrending, the first WR almost disappear (its frequency decreases by 30%) during the study period (see Figure S4). At the same time, the frequency of the fourth WR increases to more than 50%, becoming the predominant

WR (Figure S4). Thus, without detrending the four WRs are not very useful to describe atmospheric circulation in northern Chile, as most of the days belong to the same WR. However, also the overall precipitation decrease for the non-detrended wind speed fields in Figure S4 can be easily estimated by equation (2) as roughly -30% for stations to the south and along the Pacific Coast, that is a much more negative value than that of the detrended fields in Figure 8 (-10%), and more similar to the observed trend in Figure 4 (-20% to -60%, depending on the station).

Finally, it is also interesting to notice that wind anomalies of the first and fourth WRs closely resemble (albeit with a sign change in case of the fourth WR) the spatial distribution of wind trend over South America (cfr. Figure 2 with Figure 8); for this reason, their frequencies are severely influenced by the wind trend.

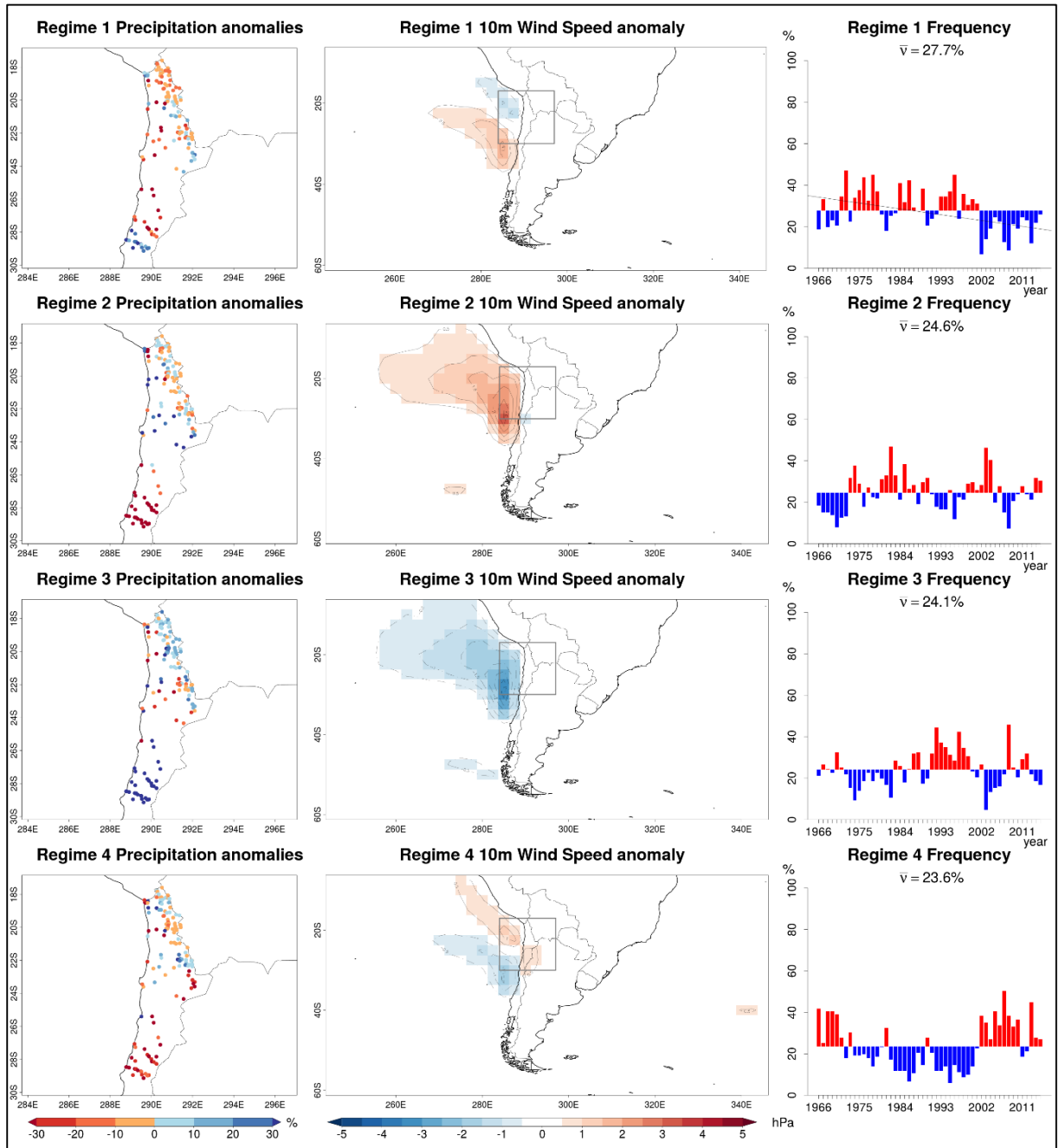

Figure 8. As Figure 5, but for 10-m wind speed anomalies (in m/s).

3.4. WR persistence and transitions

Table 1 shows the transition probabilities between the different regimes for SLP, g250, g500, and surface wind speed. At SLP, the more likely transitions follow the sequence from WR 1 to WR 4 (47.8%), then to WR 2 (50.8%), then to WR 3 (45.2%), and finally back to WR 1 (53.1%).

At the g500 level, preferred transitions are from WR 1 to WR 2 (56.7%) and from WR 2 back to WR 1 (64.5%), or from WR 3 to WR 1 (91.0%) and from WR 4 to WR 2 (87.7%).

At the g250 level, WR 1 changes to WR2 59.6% of times, while WR 2 evolves to WR 1 61.6% of times. WR shifts to WR 1 95.0% of times and WR 4 to WR 2 91.4% of times.

At the surface wind level, the preferred transitions are from WR 1 to WR 3 (42.9%), then to WR 4 (48.9%) or to WR 2 (43.9%), and finally back to WR 1 (56.8%).

Table 1. Transition probabilities (in %) for SLP, g250, g500, and wind speed levels.

| SLP transition probability (%) | | | | |
|---|---|---|---|---|
| From         To | Regime 1 | Regime 2 | Regime 3 | Regime 4 |
| Regime 1 | - | 26.2 | 25.9 | 47.8 |
| Regime 2 | 31.1 | - | 45.2 | 23.7 |
| Regime 3 | 53.1 | 39.7 | - | 7.2 |
| Regime 4 | 44.0 | 50.8 | 5.2 | - |
| g500 transition probability (%) | | | | |
| From         To | Regime 1 | Regime 2 | Regime 3 | Regime 4 |
| Regime 1 | - | 56.7 | 40.9 | 2.4 |
| Regime 2 | 64.5 | - | 5.7 | 29.8 |
| Regime 3 | 91.0 | 8.1 | - | 0.9 |
| Regime 4 | 10.8 | 87.7 | 1.5 | - |
| g250 transition probability (%) | | | | |
| From         To | Regime 1 | Regime 2 | Regime 3 | Regime 4 |
| Regime 1 | - | 59.6 | 38.7 | 1.8 |
| Regime 2 | 61.6 | - | 2.9 | 35.5 |
| Regime 3 | 95.0 | 4.5 | - | 0.5 |
| Regime 4 | 8.0 | 91.4 | 0.6 | - |
| Wind speed transition probability (%) | | | | |
| From         To | Regime 1 | Regime 2 | Regime 3 | Regime 4 |
| Regime 1 | - | 37.7 | 42.9 | 19.4 |
| Regime 2 | 56.8 | - | 6.2 | 37.0 |
| Regime 3 | 42.5 | 8.5 | - | 48.9 |
| Regime 4 | 26.3 | 43.9 | 29.9 | - |

It is interesting to discuss how initiations and terminations of WRs take place, and how long before the initiation and after the termination of each WR it is possible to identify preferred precursors and successors. The results are consistent, showing that opposite regimes (e.g., WR 3 and WR 4 for the g500 and g250 levels) have very low transition probabilities. By contrast, similar WRs (e.g., WR 1 and WR 3 for the g250 level) show high transition probabilities. Therefore, transitions exhibited between WRs are coherent with the results.

Figures 9 to 12 show the average persistence of each WR identified for SLP, 500 hPa, 250 hPa, and 10 m wind speed, respectively, with their mean annual value and significant trend (if any). Average persistence at SLP ranges from 1.9 to 2.3 days. For all WRs, persistence shows a significant trend. At the g500 level, the average persistence ranges from 2.2 days to 3.0 days, and only WR 1, with a mean persistence of 2.4 days, shows a significant trend of -0.5 days during the 50 years of the study period. At upper levels (g250), the mean persistence of the WRs ranges from 2.7 to 3.7 days, and only WR 2 shows a significant and negative trend of 0.5 days. Finally, at the 10 m wind speed level, the average persistence varies from 2.0 to 2.4 days, exhibiting a negative and significant trend in WR 1 of -0.5 days.

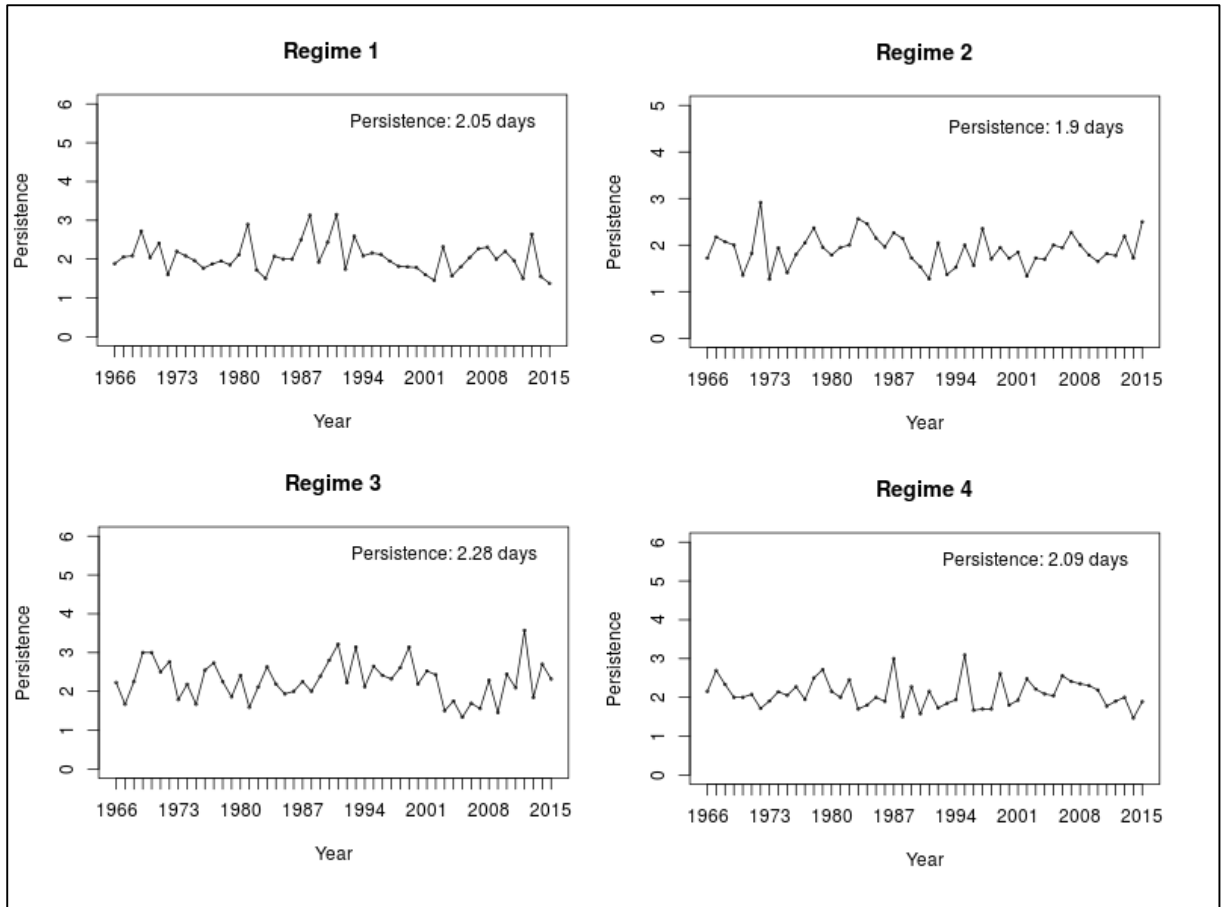

Figure 9. Inter-annual persistence (in days) for each WR at SLP level. A red line indicates a significant trend (confidence level: 95%). The average persistence value over the 1966–2015 period is shown in the top right corner.

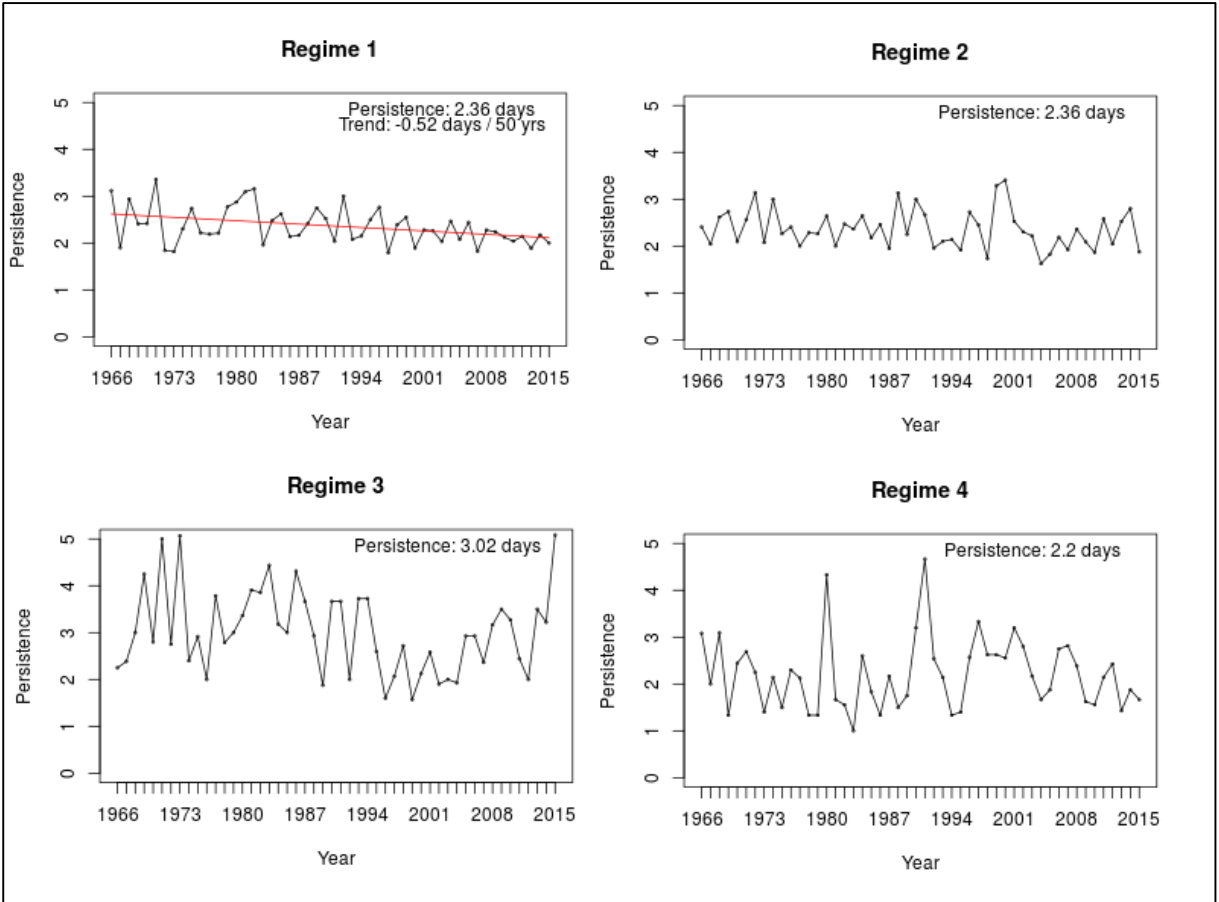

Figure 10. Inter-annual persistence (in days) for each WR at the g500 level. A red line indicates a significant trend (confidence level: 95%). The average persistence value over the 1966–2015 period is shown in the top right corner.

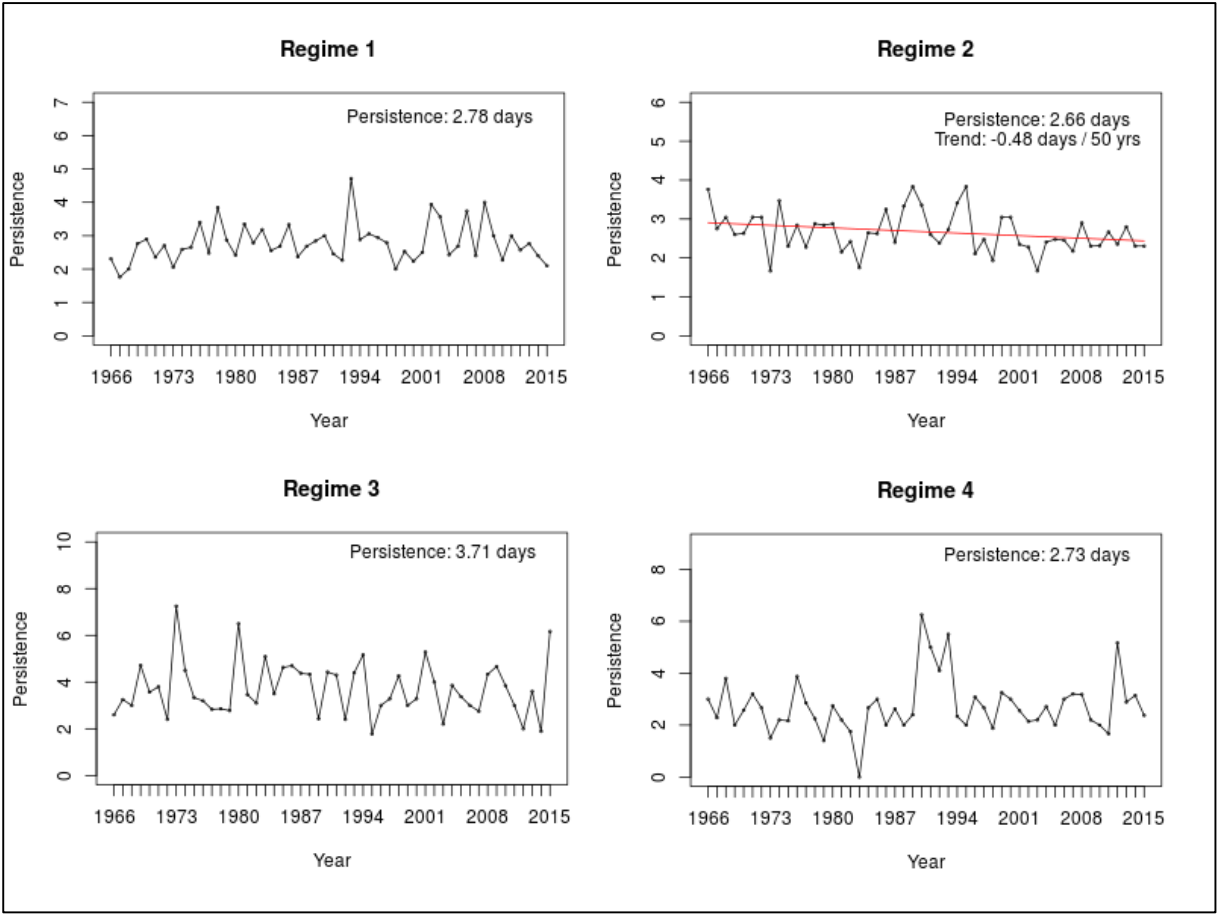

Figure 11. Inter-annual persistence (in days) for each WR at the g250 level. A red line indicates a significant trend (confidence level: 95%). The average persistence value over the 1966–2015 period is shown in the top right corner.

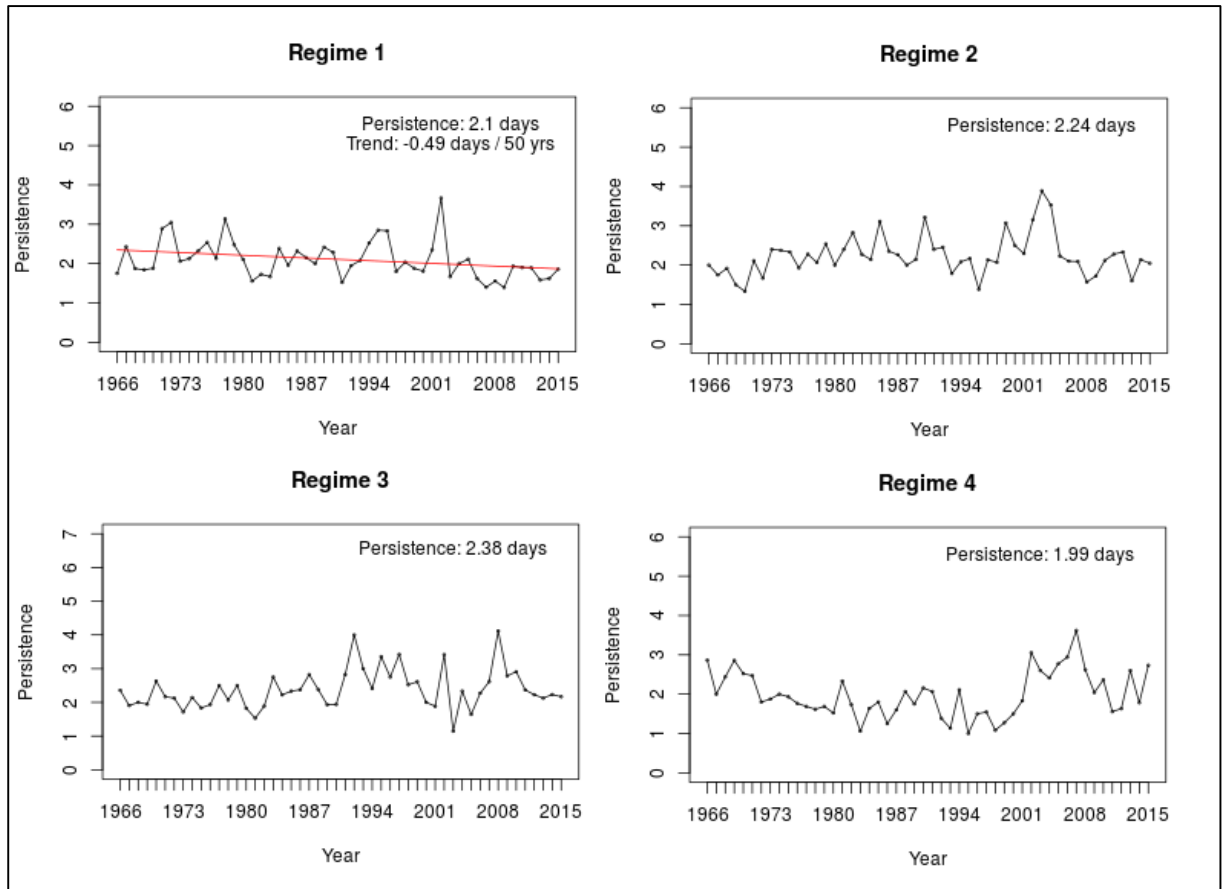

Figure 12. Inter-annual persistence (in days) for each WR at the 10 m wind speed level. A red line indicates a significant trend (confidence level: 95%). The average persistence value over the 1966–2015 period is shown in the top right corner.

WR persistence only shows three significant trends (at the confidence level of 95%), which decrease in all cases. These results are the opposite to those found for North America (Francis et al., 2018), where an increasing persistence of WRs is linked to rapid Arctic warming. The same phenomenon is not occurring in the Antarctic, therefore this could explain the general lack of trends and only three negative trends for all the WR in the Southern Hemisphere.

Lastly, it is evident that regional patterns have a direct impact on precipitation, and their future behaviour is easy to compute and simulate, even at a sub-seasonal resolution (Cortesi et al., 2019). Therefore, increasing the accuracy of WR modelling will enable more accurate precipitation modelling results to be obtained (Dawson and Palmer, 2015).

## 4. Conclusions

In this study, we classified four WRs for the wet season (December to April) between 1966 and 2015 for sea level pressure, surface wind speed, 500 hPa and 250 hPa geopotential heights. We analysed their influence on the precipitation anomalies of northern Chile.

Two of the WRs classified from surface wind speed present a long-term frequency change which determines an average decrease of rainfall amount of about -30% (in the not-detrended WR classification) in the south part of northern Chile (28-29º S), which can be compared with the intensity of the Megadrought experienced in the northern part of central Chile. Thus, these two WRs can be considered good sources of predictability of precipitation for northern Chile. Their wind patterns are also similar to those of the negative phase of the ENSO, and their frequency change determines a weakening of the subtropical westerlies over central Chile. One of these two WRs also exhibits a significant and negative persistence trend, probably due to its frequency decrease. On the contrary, WRs based on geopotential or SLP do not show any significant trend in their frequency of occurrence, even if sometimes they can highly influence local precipitation. Thus, they cannot be linked to any long-term change of rainfall amount in the study area. These WRs reproduce the situation linked to the Bolivian High (both in g500 and g250), considered as the main mechanism in the origin of precipitation in the north of the study area

(with observed positive rainfall anomalies). The opposed situation is also shown in both tropospheric levels, with negative rainfall anomalies.

Significant changes in the persistence of the different regimes over the 1966–2015 period are only found in three cases, and are linked to the lack of warming in the Antarctic as opposed to the Arctic. Preferred transitions between WRs are more likely to take place from WRs with similar patterns.

Future work could consist in applying the WR paradigm to central Chile to better focus on the Megadrought, or to employ them as a tool for evaluating climate models. Assessing the model ability to simulate observed regimes in terms of their relative frequency and spatial patterns would help provide confidence in the models. In addition, similar studies could be carried out evaluating the WR impacts on other variables such as daily temperatures, linking them to the main teleconnection patterns.

These results help to understand the mechanisms leading to precipitation in the tropical Pacific coast of South America, where water is a very scarce resource. The application of WRs to this area exhibits more accuracy than other synoptic classifications (such as the Jenkinson and Collison classifications) when working at low latitudes.

**Acknowledgements**
The authors want to thank the support of the FONDECYT Project 11160059, the Climatology Group (2017 SGR 1362, Catalan Government), the Spanish Ministry of Economy and Competitiveness (MINECO) as part of the Juan de la Cierva - Incorporación grant (BOE-A-2010-3694), the New European Wind Atlas (NEWA) project (PCIN-2016-029), the Subseasonal to Seasonal for Energy (S2S4E) project (H2020-SC5-2016-2017), and the CLICES Project (CGL2017-83866-C3-2-R). The authors also acknowledge the s2d verification R-language-based software package developers, as this package was used for the data analysis and the visualization of the results presented in this work.